# Neurophysiological Investigation of Context Modulation based on Musical Stimulus

Siddharth Mehrotra[1], Anuj Shukla[2], Dipanjan Roy[3]

*Cognitive Science Lab, International Institute of Information Technology, Hyderabad, India.*

[1]siddharth.mehrotra@research.iiit.ac.in, [2]anuj.shukla@research.iiit.ac.in, [3]dipanjan.roy@iiit.ac.in

## ABSTRACT

There are numerous studies which suggest that perhaps music is truly the language of emotions. Music seems to have an almost willful, evasive quality, defying simple explanation, and indeed requires deeper neurophysiological investigations to gain a better understanding. The current study makes an attempt in that direction to explore the effect of context on music perception. To investigate the same, we measured Galvanic Skin Responses (GSR) and self-reported emotion on 18 participants while listening to different Ragas (musical stimulus) composed of different Rasa's (emotional expression) in the different context (Neutral, Pleasant, and Unpleasant). The IAPS pictures were used to induce the emotional context in participants. Our results from this study suggest that the context can modulate emotional response in music perception but only for a shorter time scale. Interestingly, here we demonstrate by combining GSR and self-reports that this effect gradually vanishes over time and shows emotional adaptation irrespective of context. The overall findings suggest that specific context effects of music perception are transitory in nature and gets saturated on a longer time scale.

## I. INTRODUCTION

According to (Levitin, 2006), music is an organized sound, but the organization has to involve some element of the unexpected or it is emotionally flat and robotic. The appreciation we have for music is closely related to our ability to learn the underlying structure of the music we like- the equivalent to grammar in spoken or sign languages- and to be able to make predictions about what will come next. The thrills, chills and tears we experience from music are the results of having our expectations artfully manipulated by a skilled composer and the musicians who interpret the music.

There are three philosophical problems on music and emotion. The first issue concerns the fact that music is frequently perceived as a manifestation of various emotions – despite the fact that music is not a sentient being. The second issue involves the listener's emotional response to music in such instances where the listener's response mirror's the music's, despite a lack of beliefs that usually underline such a response. The third issue concerns why do a listener enjoy and revisit pieces of music that, on their own account, incline them to feel sad or happy. On examining the major philosophical theories of music such as expression theory, arousal theory, music as a virtual persona and contour theory which involves context which is attached to the music (Sharpe, 2004 & Ridley, 2004).

Music is a widely available form of media with the ability to influence attitudes and manipulate emotions (Juslin, 2008 & Wheeler, 2001) and listeners are drawn to music that reflects or improves their emotional state (Saarikallio, 2011; Thoma et al., 2012 & Papinczak et al., 2015). In Indian Classical Music, the melodic basis of compositions and improvisations is based on distinct melodic modes, called 'Ragas.' The etymological meaning of the word 'Raga' is derived from a Sanskrit word 'Ranjayati iti raga', meaning 'that which colours the mind.' Ragas are closely associated with particular emotional themes, termed as 'rasas' (emotional essence). Expression of the raga-rasa aspect is considered the primary goal in Indian Classical Music and this expression is intended to vary dynamically during a piece's performance.

The impact of music on the human body is a significant trend in music research. Different kinds of music have direct and indirect effects on physiological functions e.g. heart rate in normal and pathological conditions. Among various physiological measurements, the galvanic skin response is a noninvasive, useful, simple and reproducible method of capturing the autonomic nerve response (Vanderark, 1992). We used self-reports and GSR (Galvanic Skin Response) responses by participants as a measure to study the emotional adaptation over time while listening to ragas of Indian Classical Music. Our specific hypothesis is that adaptation will take place over time irrespective of emotional context. To tap into the adaptation effect, we will be mainly focusing on the GSR potentials and its saturation.

## II. MATERIAL AND METHODS

### A. Participants

Eighteen university students, 10 men (M = 23.3, SD = 3.66) and 8 women (M= 22.9, SD = 5.43), took part in the experiment. All participants were musically untrained and reported normal hearing. Informed consent was obtained from all participants. The Local Human Ethics Committee of the International Institute of Information Technology, India, approved the study.

### B. Apparatus

Electrodermal Activity was recorded using BIOPAC MP150 System with PPGED-R configuration. The sampling rate was 100 samples/second. EDA measurements were recorded continuously during the entire course of the experiment. The experiment was designed using PsychoPy Software (Peirce, 2007).

### C. Stimulus

*Visual Stimulus*: Visual stimuli were presented for 1.5 s in an event-related design. Three categories of stimuli were displayed from the International Affective Picture System (Lang et al., 2008). The images (pixel dimensions: 400X 300) were presented in 32-bit color (resolution: 1024X768 pixels).

i. <u>Pleasant Group</u>: 15 pleasant images were selected from IAPS images database. The valance was equal to or greater than 5 (mean pleasure rating = 7.05, SD = 0.63, range = 5.00-8.34; mean arousal rating = 4.87, SD = 0.98, range = 2.90-7.35).

ii. <u>Unpleasant Group</u>: 15 unpleasant images were selected from IAPS images database. The valance was less than the neutral midpoint of 5 (mean pleasure rating = 3.05, SD = 0.84, range = 1.45-4.59; mean arousal rating = 5.56, SD = 0.92, range = 2.63-7.35).

iii. Neutral Group: This group served as a Control Group. Subjects were shown 15 images accounting arousal and valence of range 3.5-4.5 based on IAPS normative ratings. All the IAPS pictures were classified as per recommendation of (Mikels et al., 2015).

*Musical Stimulus*: A set of 12 different ragas were taken from <u>SwarGanga Music Foundation</u>. The duration of each raga was of 3 minutes and were classified as Happy, Sad, Angry and Calm based on their rasa (see Table 1).

**Table 1.** This table lists the *ragas* used in the study and the *rasa* used by the artist to play the *raga*

| *Rasa* | *Raga* |
|---|---|
| **Hasya** (Happy) | Bhupali (Bansuri-flute) <br> Kamaj (Sitar-stringed) <br> Bhupali (Surbahar) |
| **Karuna** (Sadness) | Jogya (Dilruba-stringed) <br> Bhairavi (Bansuri-flute) <br> Bhopali-todi (Bansuri-flute) |
| **Raudra** (Anger) | Hindol (Bansuri-flute) <br> Adana (Sitar-stringed) <br> Sohini (Bansuri-flute) |
| **Shanta** (Calm) | Yaman Kalyan (Bansuri-flute) <br> Yaman (Surbahar-stringed) <br> Bhilaskhani (Dilruba-stringed) |

Therefore, in total there were 3 ragas of the same rasa. The above classification was made in accordance with previous literature by (Balkwill & Thompson, 1999; Mathur et al., 2015). The musical structure and the sequences chosen are essentially fragments of ragas. Ragas were played randomly to each participant and participant were asked to report their Mood, Emotion and Tension level on a 5 point likert scale. Mood was rated on a 1 to 5 scale where 1 being extremely annoyed, 2 being annoyed, 3 being neutral, 4 being appeased and 5 being extremely appeased. Emotion was also rated on a 1 to 5 scale where 1 being extremely sad, 2 being sad, 3 being neutral, 4 being happy and 5 being extremely happy. Lastly, for tension Level the scale was 1 being extremely tensed, 2 being tensed, 3 being neutral, 4 being relaxed and 5 being extremely relaxed.

## III. PROCEDURE

Participants were tested individually in a dimly lit soundproof room, approximately 57 cm from a 17.8' X 14.2' inches computer monitor. The loudness was set at a comfortable level by the participants at the beginning of the experiment. When the participants arrived at the laboratory, they were given a brief idea about the protocol related to the GSR recording so that the they understand the procedure and artifact can be reduced.

Participants were explicitly instructed to avoid hand movement as the electrodes were placed on index and middle finger. They were also instructed on screen which steps to follow, e.g. pressing spacebar after a visual stimulus, directions of giving ratings etc. The electrodes were attached to the participants 10-15 minutes before the experiment to record their baseline skin conductance. After receiving the instructions and having their electrodes attached, participants rated their initial mood. Participants were asked to report their emotional quotient on 3 categories namely mood, emotion, tension. Most of the participants rated their emotional quotient as calm before going on the main experiment. All the participants were randomly assigned into 3 groups namely pleasant, unpleasant and neutral and presented different emotional pictures depending on the groups e.g. in the pleasant group, participants were shown 15 pleasant picture and again asked them to rate their mood to make sure that participants should be in a particular emotional context and then they were allowed to take part in the main experiment.

After rating their emotional quotient within their respective groups, subjects were blindfolded and were given BOSE 23XSS noise cancellation headphones to listen to the musical stimuli. Ragas were presented randomly and participants provided their responses after listening to each raga for 3 minutes. We collected self-reports in between the two musical stimuli in order to minimize interference with ongoing measurement of electrodermal activity. Consequently, after the experiment, participants were debriefed about scope of experiment.

## IV. RESULTS

The analysis was performed in two steps: 1) Pre-processing of GSR Data and 2) Performing Statistical Analysis. GSR data was firstly down-sampled to 25Hz to perform adaptive smoothing. Adaptive smoothing on data was carried out to ensure that artefact removal does not cause distortions in the data. GSR data was then divided into chunks of time-blocks for each raga that was played with the use of marker embedded by PsychoPy software.

The GSR data were analysed using two-way ANOVA with the effect of context (group) on musical stimulus as between-subjects factor and effect of rasa of the raga as within-subject factors. The analysis was performed in MATLAB 2014b and SPSS version 16.0.

Results show a significant effect of group (Pleasant, Unpleasant and Neutral) in which participants were placed ($F_{(2, 60)} = 3.874$, $p = 0.026$) for first 60 seconds which is also reported in our previous findings (Mehrotra et al., *in submission*). An interaction effect was also observed among group and ragas ($F_{(6, 60)} = 2.585$, $p = 0.027$) for first 60 seconds. However, while considering 60 seconds to 120 seconds, we did not find any significance of the group in which participants were placed (F (2, 60) = 1.686, p = 0.194).

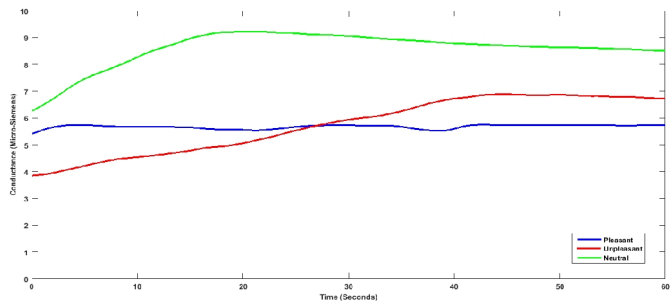

**Figure 1. GSR Potential: Angry *Raga* for *60-120 Secs* time interval**

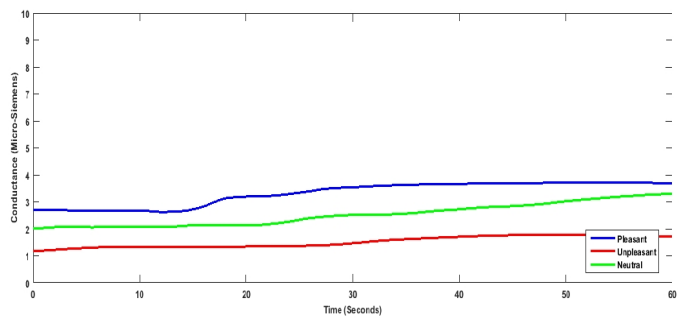

**Figure 2. GSR Potential: Calm *Raga* for *60-120 Secs* time interval**

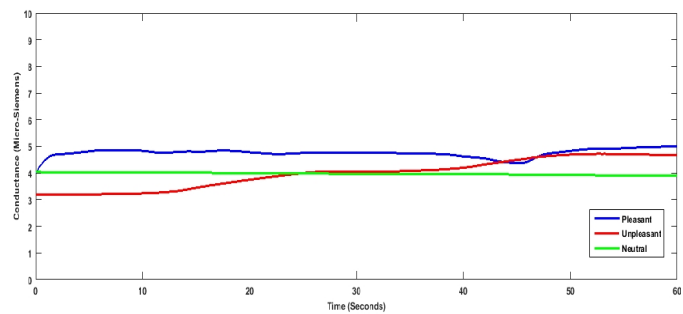

**Figure 3. GSR Potential: Happy *Raga* for *60-120 Secs* time interval**

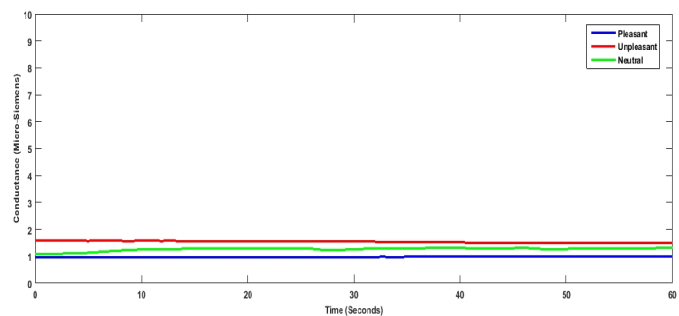

**Figure 4. GSR Potential: Sad *Raga* for *60-120 Secs* time interval**

A similar trend was observed for 120-180 seconds (F (2, 60) = 1.356, p = 0.265). In figure 2 and 3, we can see that the context has no dependence on the emotional group. For 60-120 seconds, in figure 1, Neutral group has maximum average GSR potential followed by Unpleasant and Pleasant group for angry raga but at the end of 120 secs, it appears that difference in the GSR potential between Neutral and Unpleasant has reduced and hence points towards the adaptation takes place across the group. Further, this observation compliments the statistical result. We observe a similar trend for the other ragas.

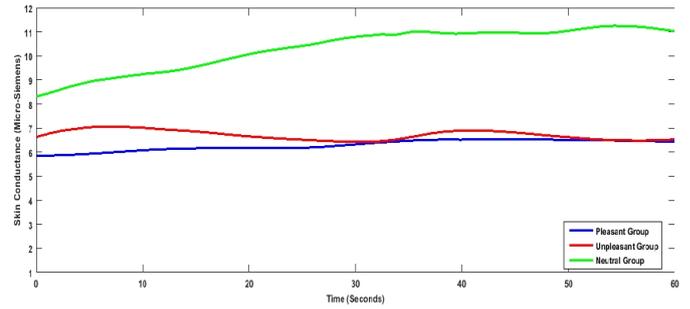

**Figure 5. GSR Potential: Angry *Raga* for *120-180 Secs* time interval**

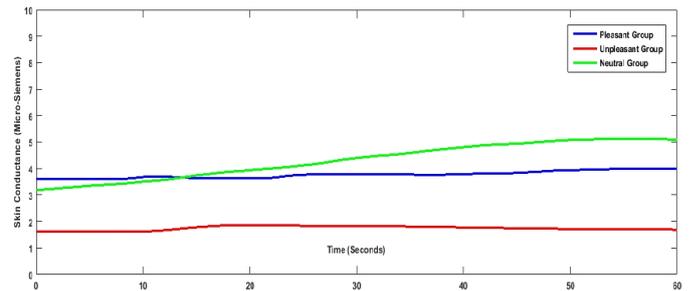

**Figure 6. GSR Potential: Calm *Raga* for *120-180 Secs* time interval**

For 120-180 seconds, the pleasant and unpleasant group have relatively same average GSR potential followed by the neutral group for angry raga in figure 5. In figure 6, from the continuation of figure 2, Neutral group has achieved highest GSR potential followed by Pleasant and Unpleasant group for Calm Raga. In figure 7, we can follow the same continuation of a trend from figure 3. The Pleasant group has a maximum GSR potential closely followed by the Unpleasant group and the minimum GSR potential is exhibited by the Neutral group for Happy raga. We interpret this as an adaption effect of the musical stimuli that is reaching a plateau over time. Adaption effect is found in both 60-120 secs and 120-180 secs for Calm, Sad, Angry and Happy raga.

In figure 8, the Unpleasant group has highest GSR potential followed by Neutral group and Pleasant group for sad raga. Sad raga shows the strongest effect of adaptation for all groups viz. the Pleasant, Unpleasant and Neutral group for the musical stimulus.

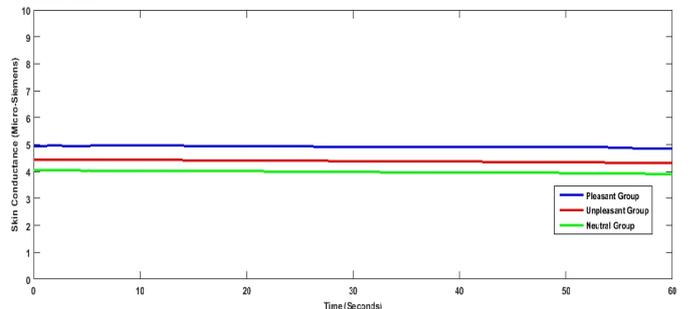

**Figure 7. GSR Potential: Happy *Raga* for *120-180 Secs* time interval**

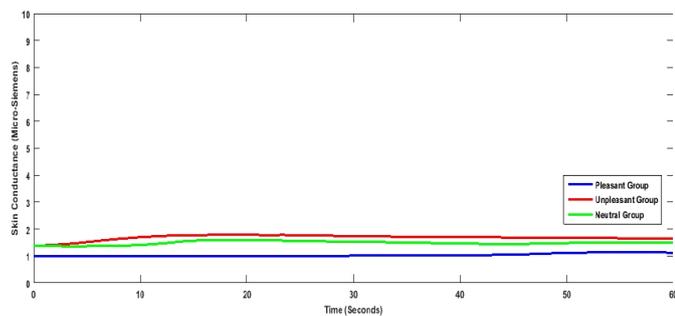

**Figure 8.** GSR Potential: Sad *Raga* for *120-180 Secs* time interval

## V. CONCLUSION

In our previous study, we have established that as the context varies, the emotion in participants also changes despite expression of the music i.e. rasa. The context plays an important role in the perception of Ragas. However, this change is temporal in nature. In the current study, our findings suggest that this effect vanishes over time and adaptation take place. We found an adaptation effect towards the musical stimulus for all participants irrespective of the group in which they are placed. The current finding provides new insight into the music perception studies and indicates that our emotional responses may not be consistent over time for the given music and might have persistence of adaptation effect. Further, the current finding needs investigation and further validation across cultures to gain more insight.

## ACKNOWLEDGMENT

We are grateful to all our participants for participating in the study. We are thankful to Prof. S.Bapi Raju for their helpful suggestions. This work was funded in part by the Kohli Center on Intelligent Systems (KCIS), as well as the International Institute of Information Technology, India.

## REFERENCES

Balkwill, L., & Thompson, W. F. (1999). A Cross-Cultural Investigation of the Perception of Emotion in Music: Psychophysical and Cultural Cues. *Music Perception: An Interdisciplinary Journal*, 17(1), 43-64

Carifio, J., & Perla, R. J. (2007). Ten Common Misunderstandings, Misconceptions, Persistent Myths and Urban Legends about Likert Scales and Likert Response Formats and their Antidotes. *Journal of Social Sciences,* 106-116. doi:10.3844/jssp.2007.106.116

Dade, P. (2010). A Dictionary of Psychology (3rd ed.)201010 Andrew M. Colman. *A Dictionary of Psychology* (3rd ed.) Oxford: Oxford University Press 2009. xi 882 pp., ISBN: 978 0 19 953406 7

Huron, D. (2003). Is Music an Evolutionary Adaptation*? The Cognitive Neuroscience of Music*, 57-75.

Lang, P.J., Bradley, M.M., & Cuthbert, B.N. (2008*). International affective picture system (IAPS)*: Affective ratings of pictures and instruction manual. Technical Report A-8. University of Florida, Gainesville, FL.

Levitin, D. (2007). *This is Your Brain on Music*. By Daniel Levitin. London: Atlantic Books, 2007. 320 pp. ISBN 978-1-84354-715-0 (hb).

Mathur, A., Vijayakumar, S. H., Chakrabarti, B., & Singh, N. C. (2015). Emotional responses to Hindustani raga music: The role of musical structure. *Frontiers in Psychology Front. Psychol., 6*. doi:10.3389/fpsyg.2015.00513

Mehrotra, S., Shukla, A. & Roy D., Effect of context on perception of *Ragas* (2016). *Psychology of Music* (submitted), SAGE publications

Mikels, J. A., Fredrickson, B. L., Larkin, G. R., Lindberg, C. M., Maglio, S. J., & Reuter-Lorenz, P. A. (2005). Emotional category data on images from the international affective picture system. *Behavior Research Methods, 37*(4), 626-630. doi:10.3758/bf03192732

Papinczak, Z. E., Dingle, G. A., Stoyanov, S., Hides, L., and Zelenko, O. (2015). Young people's use of music for well-being. *J. Youth Stud.*doi:10.1080/13676261.2015.1020935

Peirce, J. W. (2007). PsychoPy—Psychophysics software in Python. *Journal of Neuroscience Methods, 162*(1-2), 8-13. doi:10.1016/j.jneumeth.2006.11.01

Peretz, I., & Zatorre, R. J. (2003). *The cognitive neuroscience of music*. Oxford: Oxford University Press.

Saarikallio, S. (2011). Music as emotional self-regulation throughout adulthood. *Psychol.Music* 39,307–327.

Sharpe, R. A. (2004). *Philosophy of music*. Montréal: McGill-Queen's University Press.

Swar Ganga Music Foundation, http://www.swarganga.org

Thoma, M. V., Ryf, S., Mohiyeddini, C., Ehlert, U., and Nater, U. M. (2012). Emotion regulation through listening to music in everyday situation. *Cog.Emot.,*26,50

Wheeler, B. L. (n.d.). *Music therapy handbook*.Wheeler, B. L. (n.d.). *Music therapy handbook*.